
\magnification=1200
\vsize 23 true cm
\hsize 17 true cm
\baselineskip 18 pt
\centerline{{\bf THE SOURCES OF THE} { \it A} {\bf AND} {\it B} {\bf
DEGENERATE STATIC VACUUM FIELDS}}
\vskip 2 cm
\noindent
\centerline{M. A. P. Martins}
\medskip
\noindent
\centerline{Instituto de F\'\i sica}
\smallskip
\noindent
\centerline{Universidade Federal do Rio de Janeiro}
\smallskip
\centerline{C. P. 68528}
\smallskip
\noindent
\centerline{Ilha do Fund\~ao, Rio de Janeiro}
\smallskip
\noindent
\centerline{21945-970, RJ, BRAZIL}
\vskip 1 true cm
\vfill
\vskip 2cm

{\bf Abstract}- We attempt to a physical interpretation of some known
static vacuum solutions
of Einstein's equations, namely, the A and B metrics of Ehlers and Kundt. All
of them have axial symmetry, so they can be transformed to the Weyl form. In
Weyl coordinates
$\ln\sqrt{-g_{44}}$
obeys a Laplace equation, and from this a source, called {\it The Newtonian
image source} can be identified. We use the image sources to interpret the
metrics. The procedure is sucessful in some cases. In others it fails
because the Weyl transform does not have reasonable properties at infinity.
\vfill\eject
{\bf 1. Introduction}

\par
For many exact solutions of Einstein's equations the physical
interpretation is difficult and uncertain. On the simplest
non-trivial vacuum solution,  that the Schwarzschild, numerous papers
have been written and it seems to the author of this work
that there are still aspetcs of it which are puzzling. According to
Bonnor [1], general relativity cannot be understood unless one can
find the physics behind the exact solutions we know and the
essence of the interpretation of their physical meaning is to
understand their sources.
\par
In this short paper we consider the physical interpretation of
certain static and axially symmetric vacuum solutions of
Einstein's equations well known for a long time, namely, the A and B
metrics of Ehlers and Kundt. An outstanding problem with these
solutions is that, using their original coordinate, one cannot see
the nature and the and location of the sources which could give rise to these
gravitational fields. The only exception is the A1 case that
represents the Schwarzschchild solution. Hence in order to make
proposals for the material sources, one should seek a more convenient
representation for them.
\par
Following an approach presented by Bonnor [2] and also by Bonnor and
Martins [3] these solutions are transformed into the Weyl
metric. The field equations for each of the transformed metrics include a
Laplace's equation from which one can, of course, identify a
Newtonian gravitational potential. The sources of this potential have
been called [4] {\it Newtonian image sources} and we shall use them to
interpret the corresponding vacuum metrics. However, as pointed out
in [2] and [3] this procedure has to be carried out with caution
because the Newtonian potential can be misleading as a guide to the
physical meaning of relativistic sources. In all cases considered we
have found that the Newtonian potential refers to semi-infinite line
masses (silm).
Having converted tha A and B metrics to the Weyl form we give the
Newtonian image sources and these are depicted in a Figure. However in the A2
and B2 cases the interpretation of the
vacuum space-times is incomplete, as we shall show by a study of the
algebraic invariants of the Riemann tensor.
\par
For sake of completeness, in section 2 we shall sumarize some results
concerning to Weyl metrics and semi infinite line masses as presented
in [3]. In section 3 we give the Weyl transform of each of the A and B
metrics and at that stage we are ready to attempt the interpretation
of these metrics. The paper ends with a short conclusion.
\vfil\eject
{\bf 2.  Weyl Metrics and Semi Infinite Line Masses}
\bigskip
\centerline{\it{Weyl Metrics}}

\smallskip
The metrics for a static, axially symmetric vacuum fields can be
expressed in the Weyl form [5]
$$ ds^2 = e^{2\omega}\left(\;dz^2 + \;dr^2\right) +
r^2e^{-2\mu}\;d\phi^2 - e^{2\mu}\;dt^2\eqno(2.1) $$
where $\omega$ and $\mu$ are functions of $z$ and $r$.
\par
The ranges of the coordinates will be assumed to be
$$-\infty < z < \infty ,\qquad  r\geq 0 ,\qquad  0\leq \phi \leq 2\pi
 ,\qquad-\infty < t < \infty \eqno(2.2)$$
\noindent
and we shall number them
$$x^1 = z,\qquad x^2 = r,\qquad x^3 = \phi,\qquad x^4 = t$$
\noindent
points on  $\phi = 0$ and  $\phi = 2\pi$  will be indetified.
\par
Imposing the vacuum field equations  $R_{ij} = 0$  we find that  $\mu
= \sqrt {- g_{44}}$  satisfies a Laplace's equation in cilindrical
polar coordinates
$$\mu_{,11} + \mu_{,22} + {1\over r} \mu_{,2} = 0\eqno (2.3)$$
\noindent
where colon denotes ordinary partial derivative.
\par
Once $\mu$ is prescribed, $\omega$ is determinated ( up to an
arbitrary constant ) by the remaining field equations
$$\omega_{,1} + \mu_{,1} = 2 r \mu_{,1}\mu_{,2}\qquad\hbox
{and}\qquad\omega_{,2} +  \mu_{,2} = r\left(\mu^2_{,2} -
\mu^2_{,1}\right)\eqno (2.4)$$
\par
{}From the non-zero components of the Riemann tensor, for the metric
(2.1) one can form two distinct algebraic invariants which does not
vanish identically in the case of a vacuum solution, namely
$$L = R^{ijkl}R_{ijkl} = 8e^{-4\omega}\left(P^2 + Q^2 + S^2 +
2T^2\right)\eqno (2.5)$$
$$N = R^{ijkl}R_{klmn}{{R^{mn}}_{ij}} = 48e^{-6\omega}\left( Q (P S -
T^2)\right)\eqno (2.6)$$
\noindent
where
$$P = \mu_{,22} + \mu^2_{,2} + \mu_{,1}\omega_{,1} -
\mu_{,2}\omega_{,2}\eqno (2.7.a)$$
$$Q = \mu^2_{,1} + \mu^2_{,2} - {1\over r}\mu_{,2}\eqno (2.7.b)$$
$$S = \mu_{,11} + \mu^2_{,1} - \mu_{,1}\omega_{,1} +
\mu_{,2}\omega_{,2}\eqno (2.7.c)$$
$$T = \mu_{,12} + \mu_{,1}\mu_{,2} - \mu_{,2}\omega_{,1} -
\mu_{,1}\omega_{,2}\eqno (2.7.d)$$
\par
We now look for stress singularities ( sometimes called {\it conical
singularities}) on the z-axis. Such a singularity is not revealed by
an examination of L and N. A sufficient condition for the absence of
a conical singularity on the axis  r = 0 is
$$\lim_{r\rightarrow 0}\left(\mu + \omega\right) = 0\eqno (2.8)$$
\par
Notice that the invariants L and N will be singular along the parts
of the z-axis occupied by the material sources. However the Riemann
tensor components ( and so L and N, as mentioned before ) may be well
behaved even along those parts of the z-axis where (2.8) is not satisfied.
\smallskip
\centerline{\it{Semi-Infinite Line Masses}}
\smallskip
\par
For a semi-infinite line mass of density $\sigma$ per unity length,
lying on the z-axis between  $z = z_{i}$  and infinity, the Newtonian
potential is

$$\psi = \sigma \log\lbrack R_{i} + \epsilon_{i}\left( z -
z_{i}\right)\rbrack + \log C\eqno (2.9)$$
\smallskip
\noindent
where the gravitational constant is taken to unity, C is an arbitrary
constant, and
$$R_{i} = \sqrt{ \left( z - z_{i}\right)^2 + r^2 }$$
$$\epsilon_i = +1,\;\;if\;\; the\;\; silm\;\; extends\;\; to\;\; -\infty$$
$$\epsilon_i = -1,\;\; if\;\; the\;\; silm\;\; extends\;\; to\;\;  +\infty$$
\par
The general relativistic metric corresponding to this source is

$$ds^2 = X_i^{-2\sigma}\biggl[\Bigl({X_i\over
2R_i}\Bigr)^{4\sigma^2}(dz^2 + dr^2) + r^2 d\phi^2\biggr] -
X_i^{2\sigma} dt^2\eqno (2.10)$$
where
$$X_i = R_i + \epsilon_i (z - z_i)$$
\par
The quadratic invariant L and the cubic invariant N of the Riemann tensor for
the metric (2.10) are
$$L = 12\,G^2\,(1 - {2\over 3\;Y})\eqno (2.12)$$
$$N = -12\,G^3\,(1 - Y)\eqno (2.13)$$
where
$$G = 2^{2\sigma}\,C^2\sigma\,(2\sigma - 1)R^{-2(1-\sigma)}\
[csc(\theta_i/2)]^{2(4\sigma^2 - 2\sigma +1)}$$
$$Y = (1 - \sigma)(1 + 2\sigma)\,cos^2(\theta_i/2)$$
and
\noindent
${\theta_i}$ is the angle between the end of the silm ${z_i}$ and an arbitrary
point
 in the background Euclidean space with coordinates ${(z,r, \phi)}$.
Notice that the length of ${z_iP}$ is ${R_i}$.
\par
We should make some remarks about this Weyl metric before ending this
section. As it is known, in weak static fields,  $\log \sqrt{-
g_{44}}$  can be interpreted as an aproximate Newtonian potential of
the gravitational field. Thus it seems reasonable to assume that
(2.10) gives the space-time of a silm of line density $\sigma$, if
$\sigma$ is small. However:
\noindent
i) For  $\sigma = {1\over2}$  or  $\sigma = 0$,  (2.10) represents a flat
space-time.
\noindent
ii) For  $\sigma \geq 1$  it is misleading as a metric referring to a
space-time containing a single silm. If  $\sigma > 1$  it does not have
reasonable properties at infinity (because L and N diverge as
$R_i\;{\rightarrow\infty}$), indicating the presence of
addictional sources. If  $\sigma = 1$  it acquires an extra arbitrary
constant C and assumes the form
$$ds^2 = (CX_i)^{-2}
\Bigl\{(X_i/ 2R_i)^4\,( dz^2 + dr^2)
 + r^2 d\phi^2\Bigr\} + (C X_i)^2 dt^2\eqno (2.11)$$
\noindent
In [3] Bonnor interpreted (2.11) as an infinite hollow cilinder with an applied
gravitatinal field parallel to its axis.
\noindent
iii) For $ \sigma < 0$ , (2.10) apparently refers
to a silm with a negative mass density.
However when  $\sigma = - 1/2$ it is a transform of Taub`s plane metric.
\bigskip
{\bf3.  The Sources of the A and B Metrics}
\bigskip
\par
The metrics we consider in this paper and the respective ranges of their
coordinates were given
by Ehlers and Kundt [6]. We shall list them below:
$${\bf A1}:\;\;ds^2\; =\; u^2\;(dv^2\;+sin^2v\;d\phi^2)\;+\;(1\;-2m/u)^{-
1}\;du^2\;-\;(1\;-2m/u)\;dt^2\eqno (3.1)$$
where
$$0\;\leq\;v\;\leq2\pi,\;\;\;0\;<\;2m\;<\;u\;<\;\infty\;\;or\;\;2m\;<\;0\;<\;u\;<\;\infty,\eqno (3.2)$$
$${\bf A2}:\;\;ds^2\;=\;
u^2(dv^2\;+\;sinh^2v\;d\phi^2)\;+\;(2m/u\;-\;1)^{-1}du^2\;-\;(2m/u\;-\;1)dt^2\eqno (3.3)$$
where
$$0\;\leq\;v\;<\;\infty,\;\;\;\;\;\;\;0\;<\;u\;<\;2m,\eqno (3.4)$$
$${\bf
A3}:\;\;ds^2\;=\;u^2(dv^2\;+\;v^2\;d\phi^2)\;+\;u\;du^2\;-\;u^{-1}\;dt^2\eqno
(3.5)$$
where
$$0\;\leq\;v\;<\;\infty,\;\;\;\;\;\;\;0\;<\;u\;<\;\infty,\eqno (3.6)$$
$${\bf
B1}:\;\;ds^2\;=\;(1\;-\;2m/u)^{-1}\;du^2\;+\;(1\;-\;2m/u)\;d\phi^2\;+\;u^2\;(dv^2\;-\;sin^2v\;dt^2)\eqno (3.7)$$
where
$$0\;<\;v\;<\;\pi,\;\;\;0\;<\;2m\;<\;u\;<\;\infty\;\;\;or\;\;\;2m\;<\;0\;<\;u\;<\;\infty,\eqno (3.8)$$
$${\bf
B2}:\;ds^2\;=\;(2m/u\;-\;1)^{-1}\;du^2\;+\;(2m/u\;-\;1)\;d\phi^2\;+\;u^2\;(dv^2\;-\;sinh^2v\;dt^2)\eqno (3.9)$$
where
$$0\;<\;v\;<\;\infty,\;\;\;\;\;\;\;0\;<u\;<\;2m,\eqno (3.10)$$
$${\bf
B3}:\;\;ds^2\;=\;u\;du^2\;+\;u^{-1}\;d\phi^2\;+\;u^2\;(dv^2\;-\;v^2\;dt^2)\eqno
(3.11)$$
where
$$0\;<\;u\;<\infty,\;\;\;\;\;\;\;0\;<v\;<\;\infty.\eqno (3.12)$$
\noindent
In all cases  $0\;\leq\;\phi\;\leq\;2\pi$  and $0$ and $2\pi$ are indentified;
the range of t is unrestrict.
\par
It will be noticed that metrics B1, B2, B3 arise from A1, A2, A3 respectively
by the complex tranformation
$$\phi\;\rightarrow\;i\;t,\;\;\;\;\;\;\;t\;\rightarrow\;i\;\phi\eqno (3.13)$$
\par
The transformation of these metrics to hte Weyl form is straightfoward and we
outline just the case A1 as an example
. It will take the form
$$z\;=\;z(u,v),\;\;\;\;\;r\;=\;r(u,v),\;\;\;\;\;\phi\;=\;\phi,\;\;\;\;\;t\;=\;t.\eqno (3.14)$$
\noindent
Thus, the product   $g_{33}g_{44}$   is invariant and so (2.1) gives at once
$$r\;=\;u\;(1\;-\;2m/u)^{1/2}\;sin\;v\eqno (3.15)$$
\noindent
The transformation for z is got by using the fact that the Weyl sub-metric for
z and r is conformally euclidean. By inspection one finds
$$z\;=\;(u\;-\;m)\;cos\;v\eqno (3.16)$$
\noindent
The ranges (3.2) of u and v, specified for A1 yield ranges (2.2) for z and r.
\par
The use of (3.15) and (3.16) gives the {\it the Weyl of A1}, namely
$$ds^2\;=\;{(R_1+R_2+2m)^2\over 4\;R_1\;R_2}(dz^2+dr^2)+r^2\;{R_1+R_2+2m\over
R_1+R_2-2m}\;d\phi^2\;-\;{R_1+R_2-2m\over R_1+R_2+2m}\;dt^2\eqno (3.17)$$
where
$$R_1\;=\;\sqrt{\;(z+m)^2\;+\;r^2}\;\;and\;\;R_2\;=\;\sqrt{\;(z-m)^2\;+\;r^2}\eqno(3.18)$$
\par
This result is well known [7]: A1 is the Schwarzschild solution and (3.17)
is its Weyl transform.
\par
By means of some trigonometry we can show
$${[\;R_1\;-\;(z-z_1)\;]\over
[\;R_2\;-\;(z-z_2)\;]}\;=\;{[\;R_1+R_2-(z_2-z_1)\;]\over
[\;R_1+R_2+(z_2-z_1)\;]}\eqno (3.19)$$
and thus
$$\mu\;=\; 1/2\;log\;[\;R_1 -
(z+m)\;]\;\;-\;\;1/2\;log\;[\;R_2\;-\;(z-m)\;]\eqno (3.20)$$
\par
In view of (2.9) we see that the Newtonian sources of the A1 metric are two
overlaping semi-infinite rods on the z-axis, one of line density
$\sigma\;=\;+1/2$ extending from $ z_1\;=\;-m $ to $+\infty$ and the other of
line density $\sigma\;=\;-1/2$ extending from  $z_2\;=\;+m$  to  $+\infty$.
 They make up a finite rod with ends at $z_1$ and $z_2$and density
$\sigma\;=\;+1/2$ if $m\;>\;0$ and density $\sigma\;=\;-1/2$ if $\;m\;<\;0$.
The total mass of the rod is therefore m. There is no other physical
singularities on (3.17).
 The fact that the Newtonian image of the sources - represented by the
potential
$\mu\;=\;\log{\sqrt{-g_{44}}}$ in (3.17) - is a finite rod of length 2m
 (as shown in the Figure) whereas the relativistic source of A1 is of
 course, a spherical particle, serves as a warning against too facile use
of this Newtonian image.
\par
Let us turn now to the B1 metric. It is obtained from A1 by the
transformation (3.13) and it is clear that the Weyl form of B1 will be obtained
from the Weyl form of A1 by the same transformation. We therefore transform
(3.17)
by (3.13) with result
$$ds^2={(R_1+R_2+2m)^2\over4\;R_1\;R_2}(dz^2+dr^2)+{R_1+R_2-2m\over
R_1+R_2+2m}d\phi^2-r^2\Biggl({R_1+R_2+2m\over R_1+R_2-2m}\Biggr)dt^2\eqno
(3.21)$$
which, with the use of (3.19) and the identity
$$r^2\;=\;[\;R_j\;+\;(z\;-\;z_j)][\;R_j\;-\;(z\;-\;z_j)]\eqno (3.22)$$
becomes
$$\eqalignno{
ds^2&\;=\;{(R_1+R_2+2m)^2\over4\;R_1\;R_2}(dz^2+dr^2)\;+\;{r^2d\phi^2\over[R_1+(z+m)][R_2-(z-m)]}\;-\;\cr
&\;-\;[R_1+(z+m)][R_2-(z-m)]dt^2.& (3.23)\cr}$$
\noindent
Thus, for the B1 metric
$$\mu\;=\;{1\over2}\log[R_1+(z+m)]\;+\;{1\over2}\log[R_2-(z-m)]\eqno (3.24)$$
\par
The Newtonian image of the sources in this case is seen to consist of two semi
infinite line masses, each of line density $\sigma\;=\;1/2$, one extending from
$z_1\;=\;-\;m$ to ${-\;\infty}$ and the other from $z_2\;=\;+\;m$ to
${+\;\infty}$.
\par
If $m\;>\;0$ there is a empty section of the z-axis in  $-m\;<\;z\;<m$  but as
it stands the metric (3.23) does not satisfy the regularity condition (2.8)
there. This can be remedied if we introduce the transformation
$$\bar \phi\;=\;(4m)^{-1}\;\phi,\;\;\;\;\bar t\;=\;(4m)\;t\eqno (3.25)$$
\noindent
Doing this and dropping the bar over $\phi$  and  t  we shall have the Weyl
transform of the B1 metric
$$\eqalignno{
ds^2&\;=\;{(R_1+R_2+2m)^2\over4\;R_1\;R_2}(dz^2+dr^2)\;+\;{16m^2r^2d\phi^2\over[R_1+(z+m)][R_2-(z-m)]}\;-\;\cr
&\;-\;{1\over 16m^2}[R_1+(z+m)][R_2-(z-m)]\;dt^2.& (3.26)\cr}$$
\noindent
the ranges of the coordinates being as in (2.2). It applies whether m is
positive or negative, but if  $m\;<\;0$  there is no regular part of the
z-axis. (see Figure below.).
\par
In both A1 and B1 metrics The Riemann tensor invariants L and N are infinite on
the line sources and nowhere else. Moreover L and N tend to zero at coordinate
infinity away from the line sources, and the proper distances corresponding to
coordinate infinity are infinite. Thus the semi infinite line masses shown in
the Figure  are the {\it only} sources of the A1 and B1 metrics. (For a
contrasting case see below.)

\par
The Weyl transforms of the remaining A andd B metrics listed in the beginning
of this
Section can be obtained by a simple adaption of the foregoing method. We shall
give them bellow, fulfilling the regularity condition (2.8) by a transformation
of the $\phi$ and  t  coordinates wherever possible. $ R_1$  and  $R_2$  are as
defined in (3.18). In each case we state the transformation of the coordinates
required to cast each metric into its Weyl form. We shall also state the
Newtonian sources, and thes are depicted in Figure mentioned above.
\medskip
\centerline{{\it A2 metric} $(m\;>\;0)$}
\medskip
\noindent
\underbar{Weyl Form}:
$$ds^2\;=\;{[R_2-R_1+2m]^2\over 4\;R_1\;R_2}(dz^2+dr^2)\;+\;{[R_1+(z+m)]\over
[R_2-(z-m)]}r^2d\phi^2\;-{[R_2-(z-m)]\over [R_1+(z-m)]}dt^2\eqno (3.27)$$
\noindent
\underbar{Transformation required}:
$$r\;=\;u(2m/u-1)^{1/2}\;sinh\;v,\;\;\;\;z\;=\;(u-m)\;cosh\;v,\;\;\;\;\phi\;=\;\phi,\;\;\;\;t\;=\;t\eqno (3.28)$$
\noindent
\underbar{Sources}: semi infinite line mass with $\sigma\;=+\;1/2$, from
$z\;=\;+m$ to $+\infty$; semi infinite line mass with $\sigma\;=\;-1/2$, from
$z\;=\;-m$ to $-\infty$.

\noindent
The z-axis is regular on $-m\;<\;z\;<\;m$.
\medskip
\centerline{{\it B2 metric} $(m\;>\;0)$}
\medskip
\noindent
\underbar{WeylForm}:
$$\eqalignno{ ds^2 &\;=\;{[R_2-R_1+2m]^2\over
4\;R_1\;R_2}(dz^2+dr^2)\;+\;{16\;m^2\;r^2\over
[R_1+(z+m)][R_2-(z-m)]}d\phi^2\;-\;\cr
&\;-\;{1\over 16\;m^2}[R_1+(z+m)][R_2+(z-m)]\;dt^2.& (3.29)\cr}$$.
\underbar{Transformation required}:
$$r\;=\;u(2m/u-1)^{1/2}\;sinh\;v,\;\;\;\;z\;=\;(u-m)\;cosh\;v,\;\;\;\;\phi\;=\;\phi,\;\;\;\;t\;=\;t\eqno (3.30)$$
\noindent
\underbar{Sources}: semi infinite line mass with  $\sigma\;=\;1/2$  from
$z\;=\;m$  to  $-\;\infty$; semi infinite line mass with  $\sigma\;=\;+1/2$
from  $z\;=\;-m$  to  $-\;\infty$.
\par
In the A2 and B2 metrics the Kretschmann scalar L is infinite at the location
of the line sources as expected. However, one can show that L does not tend to
zero at some other regions of the coordinate infinity (e.g. as  $r\rightarrow
0$ on  $z\; =\;0$).
\noindent
Hence other sources (or fields at infinity) must be present. Thus the Newtonian
 image sources are not the only determinant of the physics of these metrics,
and
the interptetation of them by means of the Weyl transform is unsatisfactory,
or at best incomplete. In the B2 metric the image sources together constitute
in the region  $z\;<\;-m$ a semi infinite line mass with  $\sigma\;=\;1$  so
this conclusion is not surprising
in the light of Section 2. However,it is not clear why the A2 metric should not
be
determined by the Newtonian image sources.
\medskip
\centerline{{\it A3 metric}}
\medskip
\noindent
\underbar{Weyl form}:
$$ds^2\;=\;{(r+z)^2\over
4\;R}(dz^2+dr^2)\;+\;1/2\;r^2(R+z)\;d\phi^2\;-\;2\;(R+z)^{-1}\;dt^2\eqno
(3.31)$$
where
$$R\;=\;\bigl(z^2+r^2\bigr)^{1/2}\eqno (3.32)$$
\noindent
\underbar{Transformation required}:
$$r\;=\;v\;{\sqrt u},\;\;\;\;z\;=\;u\;-\;{v^2\over
4},\;\;\;\;\phi\;=\;\phi,\;\;\;\;t\;=\;t\eqno (3.33)$$
\noindent
\underbar{Source}: semi infinite line mass with  $\sigma\;=\;-\;1/2$  on
$z\;<\;0$.
\par
The positive z-axis is regular. (3.31) is, but for a trivial scale
transformation the same as (2.10) with  $\sigma\;=\;-\;1/2$,
$\epsilon\;=\;+\;1$,  $z_i\;=\;0$. It is also isometric with Taub's general
plane symmetric vacuum metric, as given in [7], eqn. (13.30).

\vfill\eject
\centerline{{\it B3 metric}}
\medskip
\noindent
\underbar{Weyl form}:
$$ds^2\;=\;{(R+z)^2\over 4\;R}(dz^2+dr^2)\;+\;{2\;r^2\over
(R+z)^2(R-z)}\;d\phi^2\;-\;{1\over2}(R+z)^2(R-z)\;dt^2\eqno (3.34)$$
\noindent
\underbar{Transformation required}:
$$r\;=\;v{\sqrt u},\;\;\;\;z\;=\;u\;-\;{v^2\over
4},\;\;\;\;\phi\;=\;\phi,\;\;\;\;t\;=\;t\eqno (3.35)$$
\noindent
\underbar{Sources}: semi infinite line mass with  $\sigma\;=\;1/2$  on
$z\;>\;0$; semi infinite line mass with  $\sigma\;=\;1$  on $z\;<\;0$.
\par
The entire z-axis is singular. The Riemann tensor invariants L and N tend
to zero at infinity except at the ends of the z-axis. Hence the semi infinite
line masses
mentioned seem to be the only sources.
\bigskip
{\bf Conclusion}
\bigskip
\par
Our attempt to interpret The A and B metrics by means of the corresponding
Weyl transforms has met with mixed success. It seems adequate for
 the interpretation of A1, B1, A3 and B3 cases. A2 and B2 are not
satisfactorily
explained by their Weyl transforms because the latter do not have reasonable
asymptotic
properties.
\par
One of the problems in interpretations metrics arises from the coordinate
freedom
inherent in General Relativity. A singularity interpreted as a semi infinite
line
mass in one system of coordinate may become quite different in another (e.g.
A3 metric (3.31) is isometric with Taub's plane-symmetric vacuum metric, as
mentioned before). Thus
we should try to find another system of coordinate in which physical
interpretation
 for both A2 and B2 might arise in a natural way. The existence of this
possible
physical interpretation will be presented in a future paper.
\bigskip
{\bf Acknowledgement}
\bigskip
\par
I wish to express my thanks to Professor W.B.Bonnor for suggesting the problem
and his
guidance to obtainning some of the results presented in this work. I also thank
Dr. M.M.Som
for reading and comments. I ackowledge the partial financial support to
CNPq of Brazil for realizing this project.
\vfill\eject
{\bf References}
\bigskip
\noindent
[1] Bonnor W B   1992   {\it Gen. Rel.Grav.}   {\bf 24}   551

\noindent
[2] Bonnor W B   1983   {\it Gen. Rel.Grav.}  {\bf 15}   535

\noindent
[3] Bonnor W B and Martins M A P   1991    {\it Class. Quantum Grav.}   {\bf 8}
  727

\noindent
[4] Letelier P S and Oliveira S R   1987   {\it J. Math. Phys.}   {\bf 28}
165

\noindent
[5] Synge J L   1960   {\it Relativity-The General Theory}  (Amsterdam:
North-Holland)  p  312

\noindent
[6] Ehlers J and Kundt W   1962   {\it Gravitation: an introduction to current
research}  ed Witten L   (John Willey: New York)

\noindent
[7] Kramer D, Stephani H, MacCallum M A H and Herlt H   1980   {\it Exact
Solutions of Einstein`s Field Equations}   (Berlin: Deutscher)

\vfill\eject
\centerline{\bf FIGURE CAPTION}
\bigskip
\par
The disposition of line sources generating the A and B metrics, according to
their Weyl transforms. In the A2 and B2 solutions additional sources seem to be
present (see text).

\bye